\documentclass[preprint,trackchanges]{aastex62}

\usepackage[version=3]{mhchem}

\turnoffeditone
\turnoffedittwo

\received{}
\revised{}
\accepted{}

\shorttitle{}
\shortauthors{Iino et al.}

\begin{document}

\title{\ce{^{14}N/^{15}N} isotopic ratio in \ce{CH3CN} of Titan's atmosphere measured with ALMA}

\author{Takahiro Iino}
\affil{Information Technology Center, The University of Tokyo, 
2-11-16 Yayoi, Bunkyo-ku, Tokyo 113-8658, Japan}
\email{iino@nagoya-u.jp}

\author{Hideo Sagawa}
\affil{Department of Science, Kyoto Sangyo University, Motoyama, Kamigamo, Kita-ku, Kyoto 603-8555, Japan}

\author{Takashi Tsukagoshi}
\affil{National Astronomical Observatory of Japan, 2-21-1 Osawa, Mitaka, Tokyo 181-8588, Japan}

\begin{abstract}
Each of the nitriles present in the atmosphere of Titan can be expected to exhibit different \ce{^{14}N/^{15}N} values depending on their production processes, primarily because of the various \ce{N2} dissociation processes induced by \edit1{different sources such as } ultraviolet radiation, magnetospheric electrons, and galactic cosmic rays.
For \ce{CH3CN}, \edit1{one } photochemical model predicted a \ce{^{14}N/^{15}N} value as 120--130 in the lower stratosphere.
This is much higher than that for \ce{HCN} and \ce{HC3N}, $\sim$67--94.
By analyzing archival data obtained by the Atacama Large Millimeter/submillimeter Array (ALMA), we successfully detected submillimeter rotational transitions of \ce{CH3C^{15}N} ($J$ = 19--18) locate at the 338 GHz band in Titan's atmospheric spectra. 
By comparing those observations with the simultaneously observed \ce{CH3CN} \edit1{($J$ = 19--18)  } lines at the \edit1{349 } GHz band, \edit1{which probe from 160 to $\sim$400 km altitude, } we then derived \ce{^{14}N/^{15}N} in \ce{CH3CN} as \edit1{125$^{+145}_{-44}$}. 
Although the range of the derived value shows insufficient accuracy due to data quality limitations, the best-fit value suggests that \ce{^{14}N/^{15}N} for \ce{CH3CN} is higher than values that have been previously observed and theoretically predicted for \ce{HCN} and \ce{HC3N}. 
This may be explained by the different \ce{N2} dissociation sources \edit1{according to the altitudes}, as suggested by a recent photochemical model. 
\end{abstract}


\section{Introduction} 
The presence of complex nitriles has been seen as one of the key features of Titan's atmosphere since a millimeter emission spectrum of the Titan's most abundant nitrile, HCN, was discovered in 1986 by the Institut de Radio Astronomie Millim$\acute{e}$trique (IRAM) using a 30-m single-dish telescope \citep{Paubert1987}. Since that time, other nitriles such as \ce{HC3N}, \ce{CH3CN}, HNC, and \ce{C2H5CN} have been discovered by space and ground-based millimeter and \edit1{submillimeter} observations (\cite{Marten2002}; \cite{Moreno2011}; \cite{Cordiner2015}). It is currently thought that the production of the above-described nitriles starts with the dissociation of \ce{N2} into \ce{N}-atoms, which is induced by ultraviolet (UV) radiation, magnetospheric electrons (ME), and galactic cosmic rays (GCR). 
Recently, \cite{Loison2015} developed a neutral and ion chemistry model with which they successfully reproduced the observed vertical profiles of nitriles. 
According to their model, the main production path of \ce{HCN} is as follows: \ce{N(^{4}S) + CH2 -> H2CN}, \ce{H2CN + H -> HCN + H2}, where \ce{N(^{4}S)} is a ground-state \ce{N}-atom. 
In turn, \ce{N(^{2}D)}, an excited-state \ce{N}-atom, easily reacts with \ce{CH4}, which is the second most abundant species on Titan, and produces \ce{CH2NH}. \ce{CH2NH} is lost by photo-dissociation and produces \ce{H2CN}, which leads to \ce{HCN} production. Once \ce{HCN} is formed, its photolysis produces reactive \ce{CN} radicals, which subsequently react with \ce{C2H2} to produce \ce{HC3N}. Here, it is noted that the production process for \ce{HNC}, which is an isomer of \ce{HCN}, is similar to that for \ce{HCN}. \ce{HNC} is a minor product of the \ce{H2CN + H} reaction.

The production path for \ce{CH3CN} is expected to be different from that for \ce{HCN} and its daughter species. \cite{Loison2015} suggested that the main production path for \ce{CH3CN} is: \ce{N(^{4}S) + C2H3 -> CH2CN}, \ce{CH2CN + H + M -> CH3CN}. \ce{CH2CN} produces \ce{C2H5CN} by reacting with abundant \ce{CH3} radicals, and it is expected that the production of \ce{C2H3} radical occurs only in high-pressure (low altitude) regions \citep{Loison2015}. At such low altitudes, photo-dissociation of \ce{N2} hardly ever occurs. Instead, it is considered likely that the production of \ce{N(^{4}S)} at those altitudes can be attributed to GCR-induced \ce{N2} dissociation. A recent model by \cite{Dobrijevic2018} showed that, below 600 km, the production of \ce{N(^{4}S)} is dominated by GCR-induced dissociation, not by UV or ME. Thus, the production of \ce{CH3CN} and \ce{C2H5CN} can be expected to continue below 600 km. 

As mentioned above, the different \ce{N2} dissociation processes lead to complex nitrogen compositions and chemistry. 
This is important because observations of the nitrogen isotopic ratio \ce{^{14}N/^{15}N} can be of use when evaluating theoretical models of Titan's atmospheric chemistry. 
\cite{Dobrijevic2018} predicted that the \ce{^{14}N/^{15}N} measured for the nitriles would change significantly depending on their production processes, in particular the dissociation process of \ce{N2}. 
Here, it should be noted that although dissociation due to GCR does not produce any isotopic fractionation, the same is not true for dissociation due to UV and ME. 
Due to the self-shielding nature of UV, \ce{N^{14}N} is not dissociated effectively in the upper stratosphere, whereas ME dissociate \ce{N^{14}N} $\sim$10 times more effectively than UV \citep{Dobrijevic2018}. 
In turn, in case of \ce{N^{15}N}, UV dissociation is stronger than that of ME because of less self-shielding. 

The photochemical model of \cite{Dobrijevic2018} suggested that \ce{CH3CN} and \ce{C2H5CN} have large differences in \ce{^{14}N/^{15}N} values compared with those for \ce{HCN} and \ce{HC3N} because of their production path differences.
More specifically, the modeled \ce{^{14}N/^{15}N} values for \ce{CH3CN} and \ce{C2H5CN} are $\sim$80 at high altitudes ($\sim$800 km) and increase to $\sim$120 in the lower stratosphere ($\sim$200 km), while those for \ce{HCN} and \ce{HC3N} are expected to be relatively constant at $\sim$80 altitudes below 800 km. 
These modeled values for \ce{HCN} and \ce{HC3N} are in good agreement with previous observation results of 94$\pm$13 \citep{Gurwell2004} and 72.2$\pm$2.2 \citep{Molter2016} for \ce{HCN}, and 67$\pm$14 \citep{Cordiner2018} for \ce{HC3N}. 
\edit1{Interestingly, there is another photochemical model showing different \ce{^{14}N/^{15}N} values for nitriles. 
\cite{Vuitton2019} modeled nitrile chemistry including isotopic fractionation processes, which indicated that \ce{^{14}N/^{15}N} for \ce{CH3CN} at high ($>$1000 km) altitude is smaller than that of HCN and \ce{HC3N} because they proposed that \ce{CH3CN} is produced from N($^2$D) which is enriched in \ce{^{15}N}. 
In turn, in the lower stratosphere (at 200 km), three nitriles are expected to exhibit similar \ce{^{14}N/^{15}N} values of $\sim$55 because non-fractionated N-atoms produced by GCR collision with \ce{N2} homogenizes \ce{^{14}N/^{15}N} in nitriles via recycling processes. 
}

\edit1{
As mentioned above, GCR influx is a possible candidate that affects the trace species composition on Titan. 
To constrain and evaluate the effect of GCR influx, observational derivations of \ce{^{14}N/^{15}N} for \ce{CH3CN} are crucial. 
}
\edit1{Recently, \cite{Palmer2017} detected a \ce{CH3C^{15}N} signal in the submillimeter spectra taken by the Atacama Large Millimeter/submillimeter Array (ALMA), and reported \ce{^{14}N/^{15}N} in \ce{CH3CN} as 89$\pm$5.
In the analysis, the \ce{CH3CN} abundance profile was assumed to be the same as a reference profile derived in \cite{Marten2002}, which means that the respective observations of \ce{CH3C^{15}N} and \ce{CH3CN} were conducted at different respective epochs of 2014 and 1998, with significantly different spatial resolutions. 
In addition, while the detail of the observation has not yet been published in article, \edit2{\cite{Cordiner2015b}} also analyzed ALMA data and reported a low \ce{^{14}N/^{15}N} value in \ce{CH3CN} as 58$\pm$8, which is consistent with that modeled by \cite{Vuitton2019}.}

It should be noted that a similar GCR-induced trace species production and brightness change process have also been proposed for Neptune (\cite{Lellouch1994}; \cite{Aplin2016}). 
From the viewpoint of planetary science, an understanding of the effects of GCR on atmospheric compositions is important because the process may be universally applicable to planets. 

\edit1{
In this study, we attempt to measure \ce{^{14}N/^{15}N} value for \ce{CH3CN} from simultaneously obtained \ce{CH3CN} and \ce{CH3C^{15}N} data. 
Both the atmospheric structure and the abundances of nitriles on Titan, including \ce{CH3CN}, are known to exhibit seasonal and spatial variation.
In addition, due to the limited coverage of the data sampling in \edit2{$uv$} plane, the synthesized beam of the radio interferometer has elliptical shape and varies with time of the observations. 
Therefore, for the precise determination of \ce{^{14}N/^{15}N} in \ce{CH3CN}, both isotopolologues are to be recorded quasi-simultaneously with a common synthesized beam. 
}

\section{Data analysis}
\subsection{Extracting Titan data from the ALMA archive}
The atmosphere of Titan is known to have seasonal variations in its thermal structure (e.g., \cite{Achterberg2008}; \cite{Vinatier2010}; \cite{Coustenis2013}; \cite{Coustenis2018}; \cite{Thelen2018}). 
The distribution of \ce{CH3CN} within Titan's atmosphere also shows strong seasonal variations and spatial inhomogeneity (e.g., \cite{Cordiner2015}; \cite{Lai2017}; \cite{Thelen2019}).
Therefore, researchers must always be careful when interpreting an isotopic ratio if the values were derived from the data of two isotopologues taken at different observation dates, and/or with different spatial resolutions. 
Keeping this point in mind, we searched the publicly released ALMA archive to identify the most favorable data for analyzing \ce{CH3CN} and \ce{CH3C^{15}N}. 

\ce{CH3CN} and \ce{CH3C^{15}N} have a large number of spectral lines with a variety of line strength in millimeter and \edit1{submillimeter} wavelength. 
Data in the ALMA archive contain Titan as a science target and also as a flux calibration data.
Accordingly, we began with a data search conducted among spectral cubes (spatial maps having spectral dimensions). 
However, since spectral cubes for use as calibration data had not previously been prepared in the archive, it was necessary for us to reduce them from raw visibility data. 
The calibration was done using the QA-2 pipeline script \edit1{produced by the Analysis Utilities given by the ALMA observatory}. 
\edit1{In addition, we have removed spectral line flagging process in the script to preserve Titan's atmospheric spectra.}
Calibrated datasets for Titan were split into each spectral windows (SPW), and their line-of-sight Doppler-shift velocity values were corrected. 
The continuum and line emissions were separated in the $uv$ plane.
This process was needed to improve the dynamic range of the synthesized images, as our target spectral line features have weak intensity values. 
At the same time, synthesis images including the continuum emission are also required since we need to calibrate the flux density of Titan with respect to simulated spectra from our radiative transfer model (see Section 2.2). 
Imaging was performed with the CLEAN method with $csclean$ mode to produce a 320$\times$320 pixel image with a 0.025$\arcsec$ spacing. 
A CLEAN mask with 1.5$\arcsec$ diameter circle was placed at the Titan's disk center.

By examining all the CLEANed Titan images in the ALMA archival data, we successfully detected spectral features of \ce{CH3C^{15}N} in several datasets. 
We found that the data of ObsID 2013.1.00033.S contained two SPWs that include rotational transitions of \ce{CH3C^{15}N} ($J$=19--18) in the 338.88--338.92 GHz range and \edit1{\ce{CH3CN} ($J$=19--18) in the 349.21--349.45 GHz range, respectively, in a same observation performed on 29 April 2015.}
\edit2{Note that both data were obtained for the flux calibration purpose, and \ce{CH3CN} ($J$=19--18) was flagged out in the originally archived data for the precise measurement of Titan's continuum emission.}
The observation time was 157 seconds with seven observation scans taken at 6.05 second intervals.
The observations were performed for flux and amplitude calibration. 
The center frequencies of SPWs for \ce{CH3C^{15}N} and \ce{CH3CN} were 337.995 and \edit1{350.116} GHz, respectively. 
Since the total bandwidth and number of channels were 1874.956 MHz and 3840 channels, respectively, the spectral resolution was 488.270 kHz.
The number of 12-m antennas used for the observation was 39.
\edit1{The quasar J2056-4714 was used for bandpass and phase calibration}. 
The synthesis beam size was 1.19$\arcsec \times$0.62$\arcsec$ (position angle = $-$70.3$^{\circ}$). 
The apparent disk diameter (2575 km) of Titan was 0.78$\arcsec$ at the observed time, and the sub-Earth latitude was 24.7$^{\circ}$N. 

\begin{figure}
\begin{center}
\includegraphics[scale=0.4]{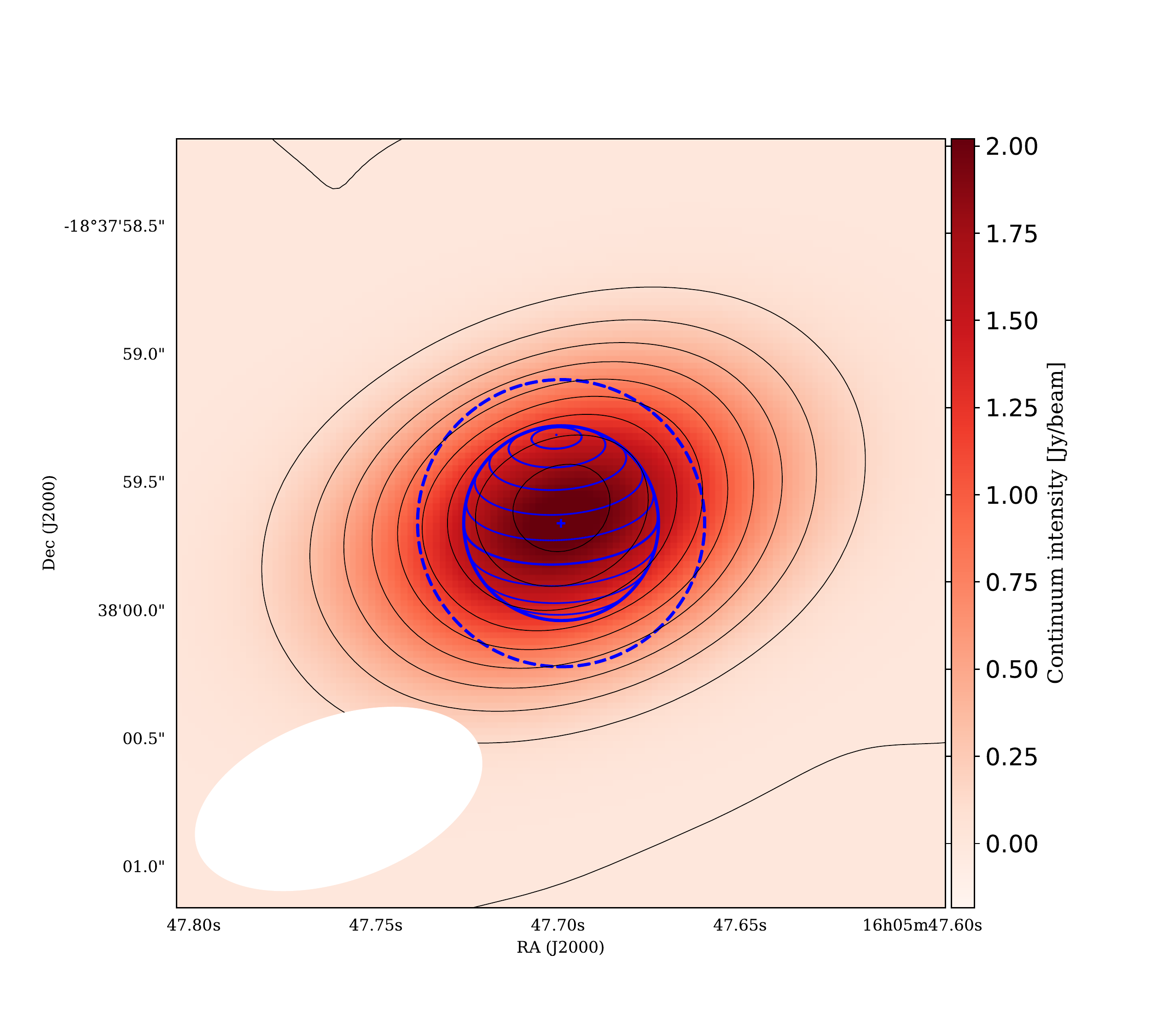}
\caption{A map of continuum (color scale) and \ce{CH3CN} integrated intensity (contours). \edit1{Contours are plotted with 0.1 Jy/beam$\cdot$km/s step.} An ellipse at the bottom left indicates the synthesized beam size. The blue cross marker is the center position of Titan. The blue dashed circle has a 1.15$\arcsec$ diameter (1200 km above the surface), an area in which the disk-averaged observed spectra were collected. \edit1{Latitudinal lines are plotted with a 15$^{\circ}$ step. }
}
\label{fig:CH3CN_map}
\end{center}
\end{figure}

Figure \ref{fig:CH3CN_map} shows an integrated intensity map for \edit1{a \ce{CH3CN} line of $J$ = 19(3) -- 18(3) which is the strongest among other lines in the $J$ = 19 -- 18 band. }
Here it can be seen that the intensity peak is located in the northern mid-latitude region, which is consistent with past observations of the seasonal behavior of this species \citep{Thelen2019}. 
The thick Titan atmosphere creates significant limb brightening, and the emission originally extends outside of the 0.78$\arcsec$-diameter disk. 
\edit1{Moreover, the limb emission is spatially spread due to the synthesized beam.}
In order to collect such an extended emission, we averaged the spectra within a region that included the Titan atmosphere up to an altitude of 1200 km ($\sim$1.15$\arcsec$ diameter disk area, which is expressed as a dashed line in Figure \ref{fig:CH3CN_map}). 
The disk-averaged spectra of both isotopologues are displayed in Figure \ref{fig:CH3CN}(a) and \ref{fig:CH3C15N}(a). 

\subsection{Radiative transfer analysis}
\begin{figure}
\begin{center}
\includegraphics[scale=0.6]{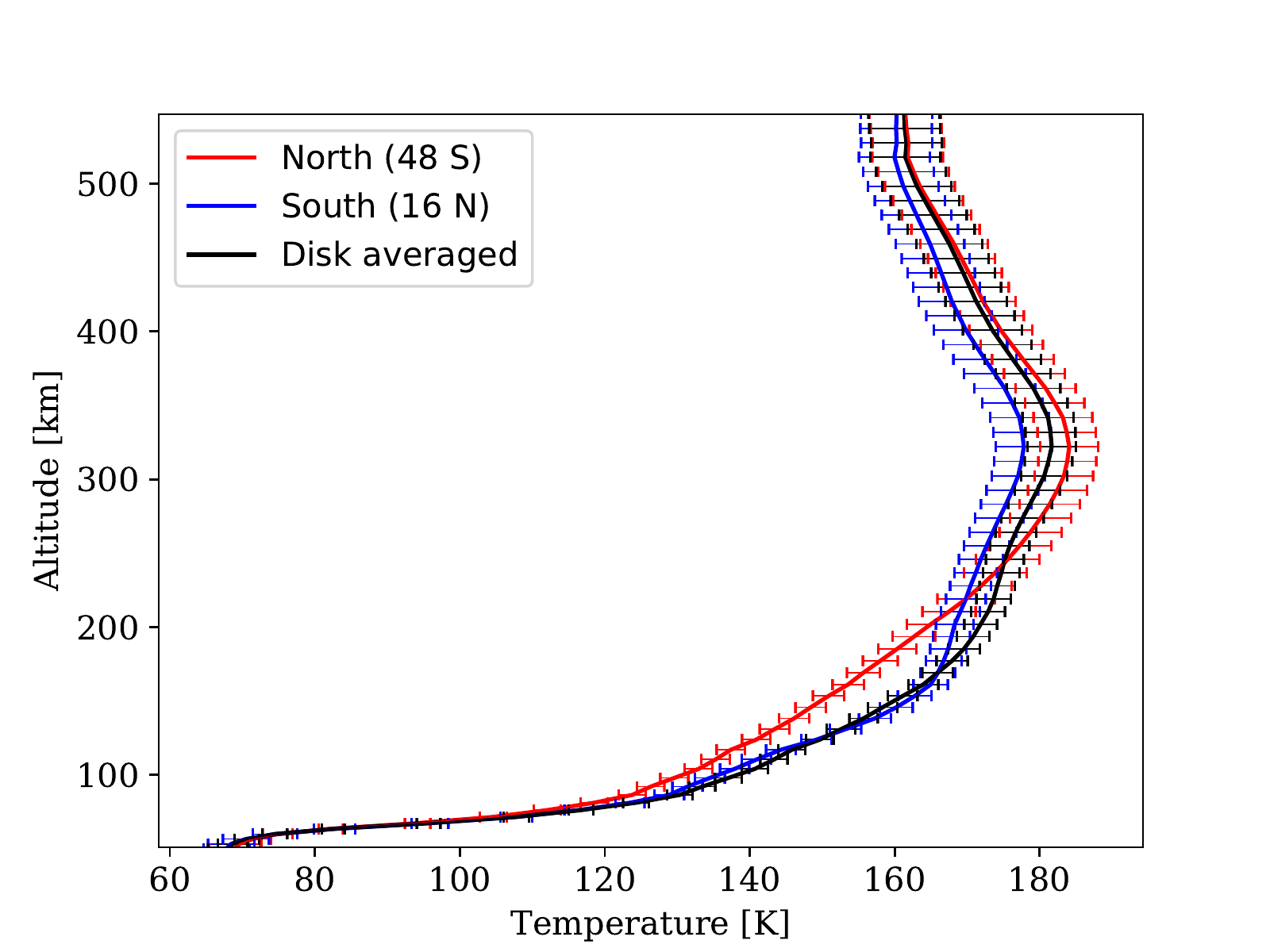}
\caption{The temperature profiles used in this study, based on the work by \cite{Thelen2018}. 
The disk-averaged profile is used as the reference profile. 
The northern 
hemisphere profile is used for the error analysis described in Section 3.2.
}
\label{fig:temperature}
\end{center}
\end{figure}

Next, we calculated the radiative transfer of atmospheric emissions from Titan. 
Our radiative transfer model \edit1{is based on the one used in \cite{Rengel2014}, which} consists of a line-by-line calculation using spherically uniform atmospheric layers. \edit1{The atmosphere from the surface up to 900\,km is divided into 240 layers (2$-$10 km depth).} 
For the temperature profile, we referred to a recent work by \cite{Thelen2018}, in which a disk-averaged temperature profile was retrieved from ALMA archived data observed on June 27, 2015 ($\sim$2 months' difference from the observation date of the data used in this study) (Figure \ref{fig:temperature}). 
An a priori profile for \ce{CH3CN} was taken from \cite{Marten2002}.
Pencil-beam synthesis spectra were calculated under various emission angle conditions, including the limb-viewing geometry, and then convolved with an elliptic beam of the CLEANed image. 
The flux density for the observed spectrum was then calibrated against the continuum brightness of the forward model spectrum. 

\begin{figure}
\begin{center}
\includegraphics[scale=0.6]{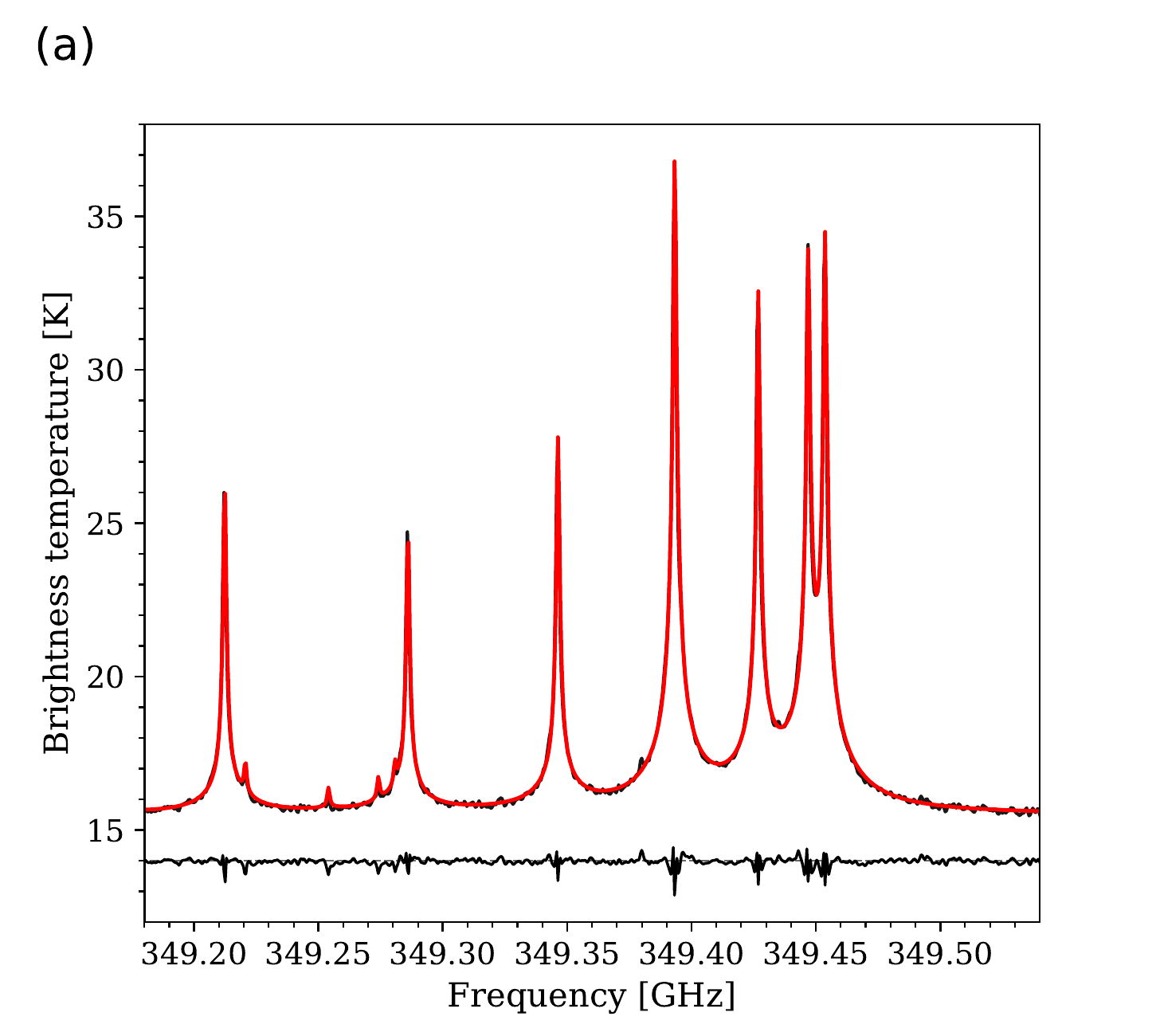}
\includegraphics[scale=0.6]{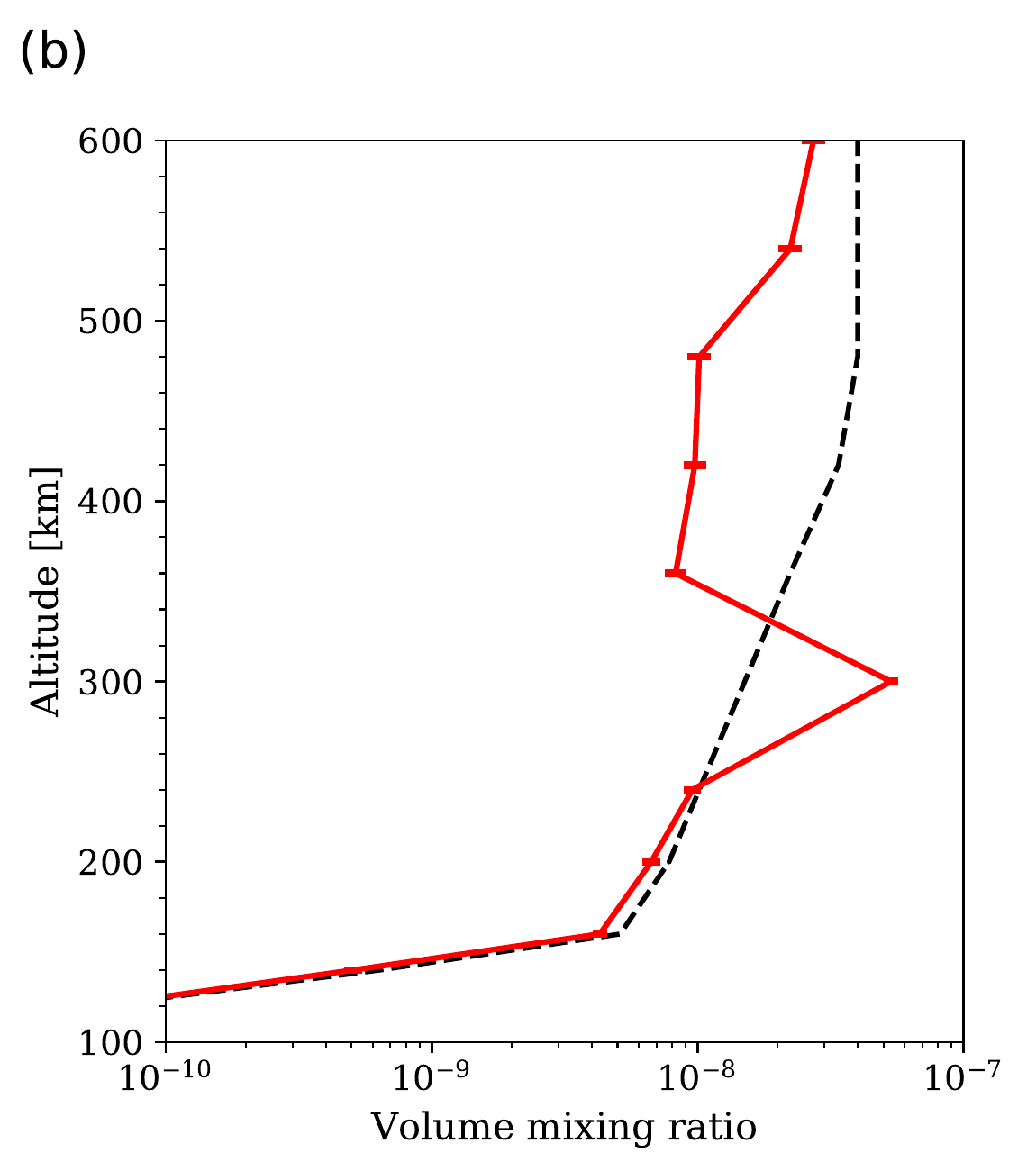}
\caption{(a) The observed (black) and best-fit (red) spectra of \ce{CH3CN}. The corresponding residual (observation - model) is plotted in the bottom of the panel with an offset of 14. 
(b) The a priori (black dashed) and retrieved (red) vertical abundance profile of \ce{CH3CN}. 
}
\label{fig:CH3CN}
\end{center}
\end{figure}

A vertical profile of \ce{CH3CN} was retrieved from the \edit1{$J$ = 19--18} transitions at 349.21--349.45 GHz. 
Figure \ref{fig:CH3CN}(a) shows the observed and best-fit spectra of \ce{CH3CN}. 
The a priori and retrieved vertical profiles are shown in Figure \ref{fig:CH3CN}(b).
Since the goal of this paper is to discuss \ce{^{14}N}/\ce{^{15}N} value for \ce{CH3CN}, we will not discuss the result of the retrieved \ce{CH3CN} profile in detail, but we briefly compared it with the work by \cite{Thelen2019} which has retrieved \ce{CH3CN} profile from the data taken on May 19, 2015. 
\edit1{Our disk-averaged \ce{CH3CN} profile shows an increase of the volume mixing ratio at the middle stratosphere ($\sim$300 km) and a decrease at $\sim$360 -- 420 km. 
Such a vertical oscillation is also seen (although not fully consistent) in the result of the northern hemispheric \ce{CH3CN} profile in \cite{Thelen2019}. }

\begin{figure}
\begin{center}
\includegraphics[scale=0.6]{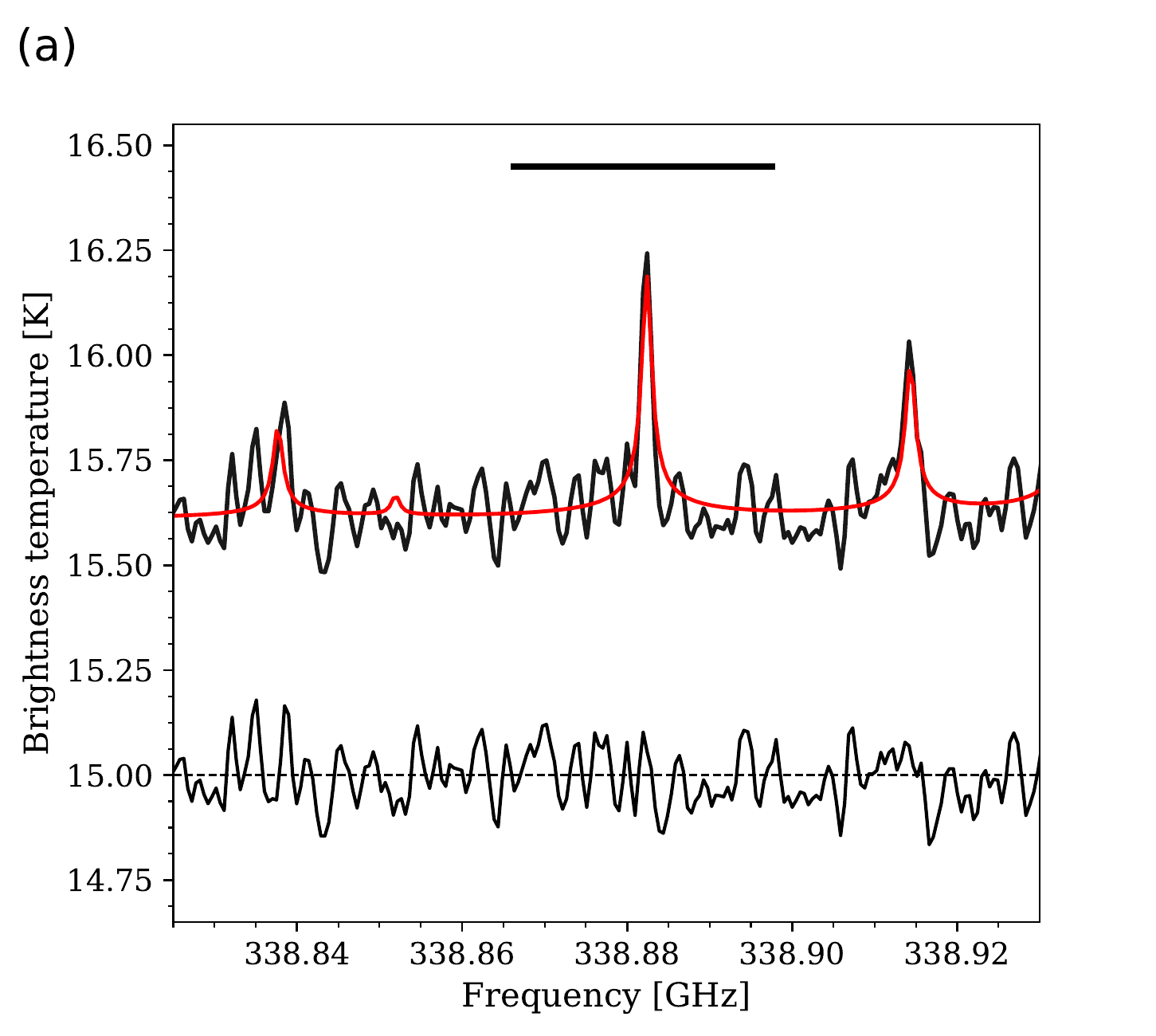}
\includegraphics[scale=0.6]{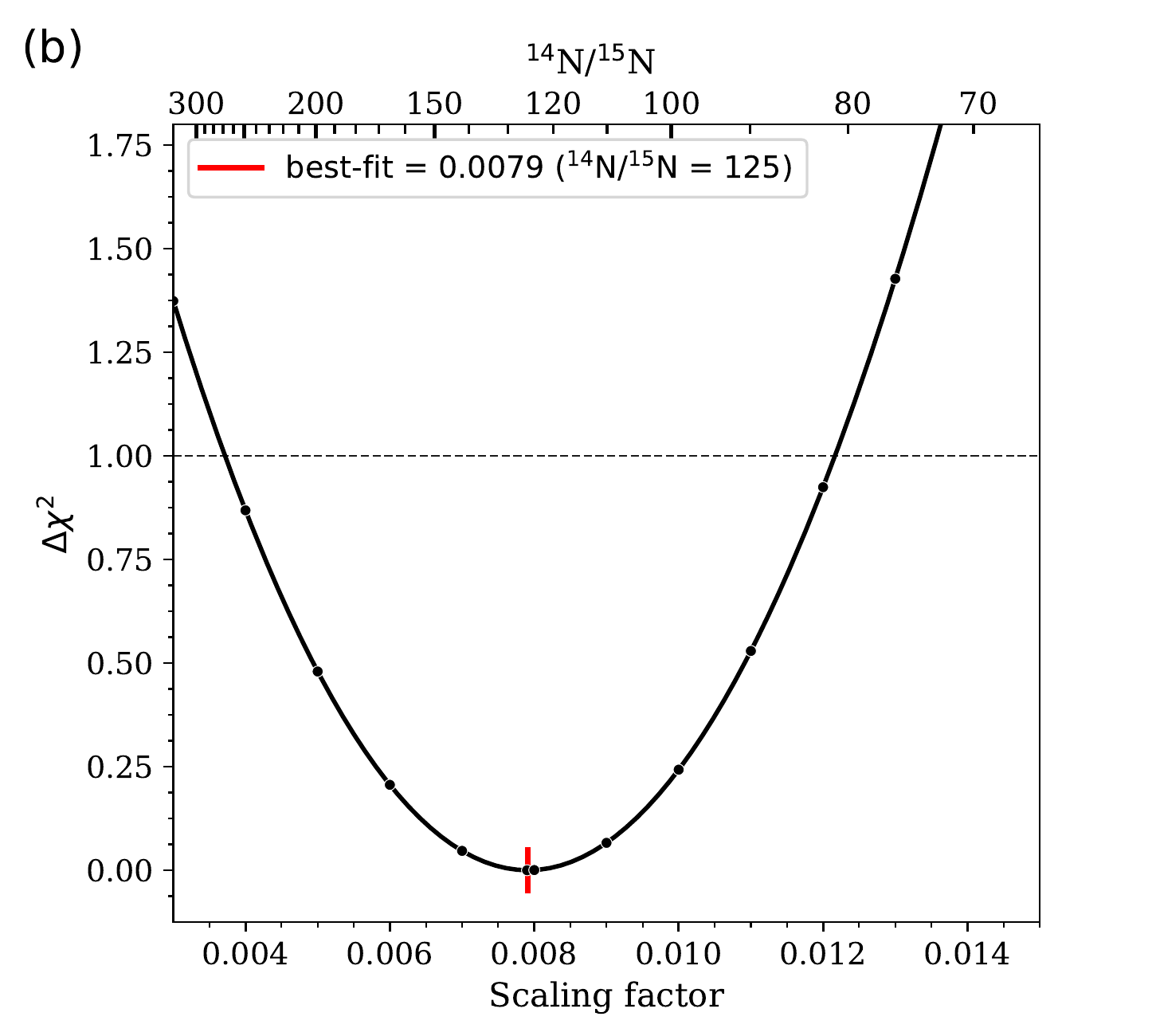}
\caption{(a) The observed (black) and best-fit (red) spectra of \ce{CH3C^{15}N}. The residual is presented with an offset of 15. The horizontal bar above the spectra indicates the spectral range used in the $\chi^2$ calculation. 
(b) $\Delta \chi^2$ values as a function of \edit1{the scaling factor of \ce{CH3CN} profile}. 
Corresponding \ce{^{14}N/^{15}N} values are shown in the top horizontal axis.}
\label{fig:CH3C15N}
\end{center}
\end{figure}

The retrieved \ce{CH3CN} profile was used to constrain the \ce{CH3C^{15}N} abundance by fitting the observed spectrum with a forward model spectrum. 
A vertically constant isotopic ratio is assumed in this study. 
The reduced $\chi^{2}$ values were calculated as a function of the scaling factor of \ce{CH3C^{15}N} with respect to the \ce{CH3CN} volume mixing ratio. 
Figure \ref{fig:CH3C15N}(a) shows the best-fit spectrum. Three \ce{CH3C^{15}N} lines are clearly detected in the observed SPW, and $\chi^{2}$ values were calculated using the data within a spectral range of $\pm$16 MHz around the best signal-to-noise ratio line at 338.882 GHz. 
The best-fit (the minimum of $\chi^{2}$) was obtained when the \edit1{scaling factor} was \edit1{0.0079}, which corresponds to the \ce{^{14}N/^{15}N} value of \edit1{125}.
It is noted that this isotopic ratio also reproduces the other two \ce{CH3C^{15}N} lines that are not used in the fitting analysis.  
The fitting uncertainty was estimated from the range of $\Delta \chi^{2} \leq 1.0$, where $\Delta \chi^{2}$ is a difference from the minimum $\chi^{2}$ value. 
The corresponding uncertainty range was obtained as \edit1{\ce{^{14}N/^{15}N} = 81--270} (Figure \ref{fig:CH3C15N}(b)). 
\edit1{The derived error range includes the previously reported \ce{^{14}N/^{15}N} value of 89$\pm$5 \citep{Palmer2017}. }

\section{Discussion}
\subsection{Caveat of the data analysis}
\begin{figure}
\begin{center}
\includegraphics[scale=0.605]{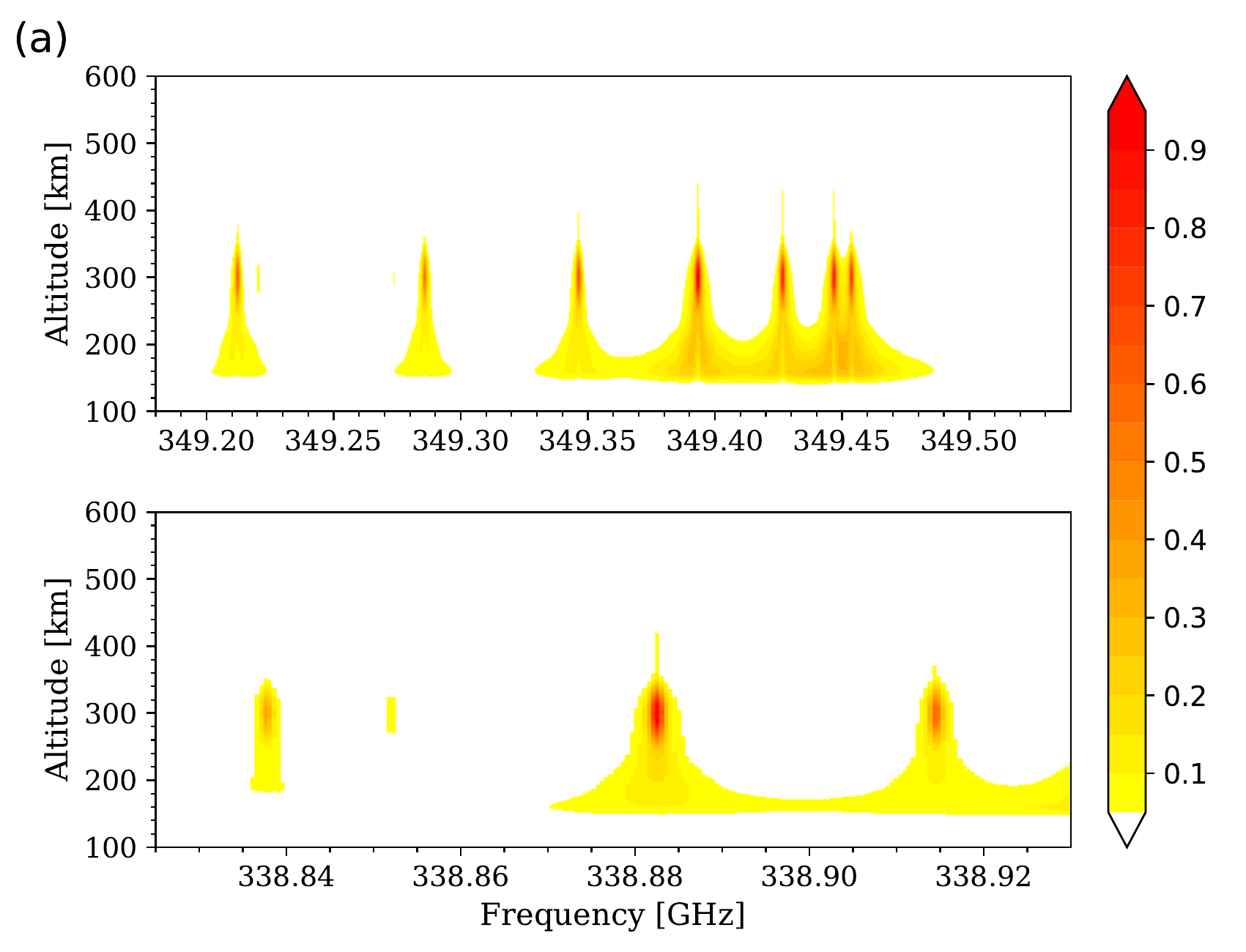}
\includegraphics[scale=0.605]{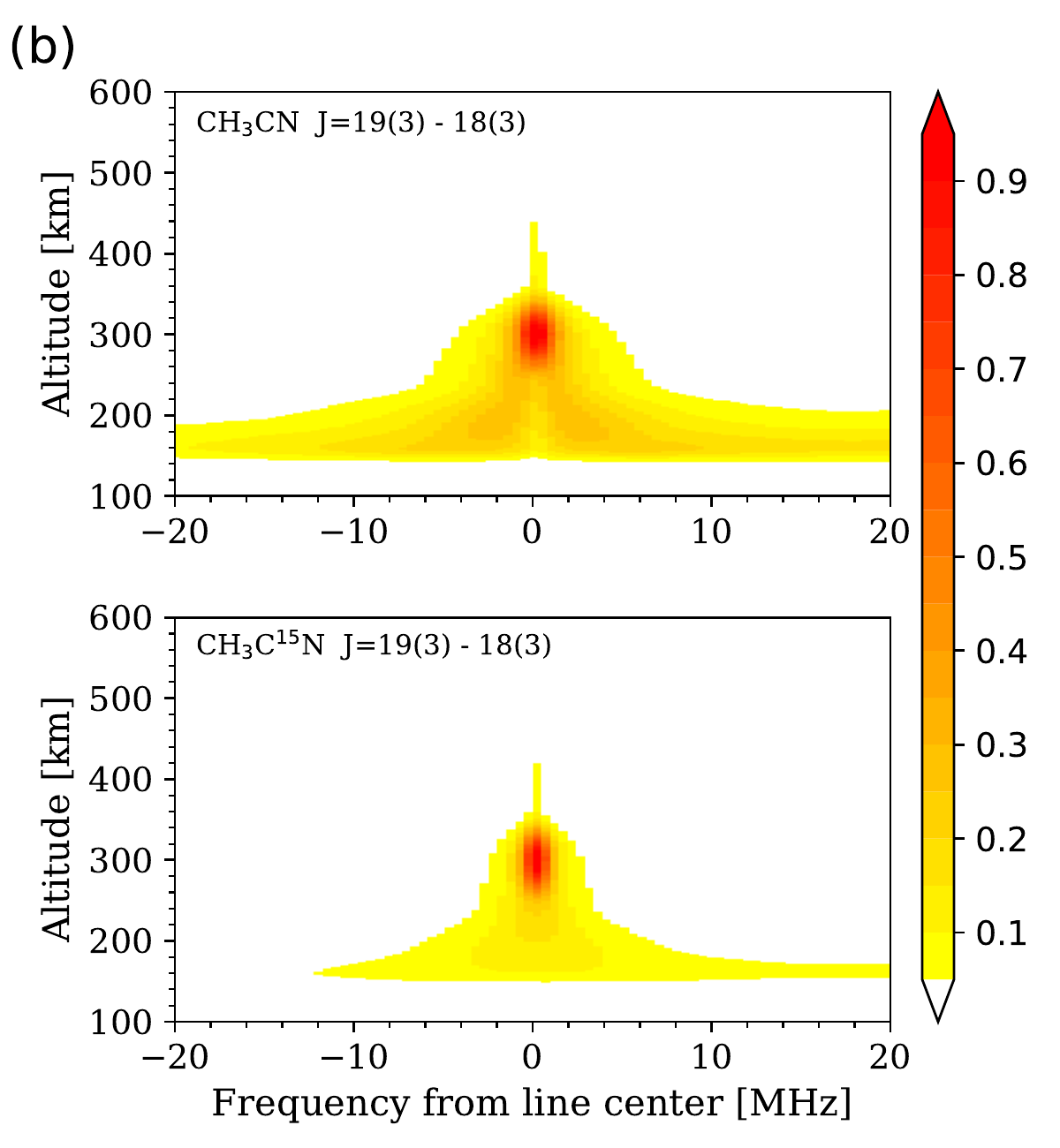}
\caption{\edit1{(a) Color maps of normalized Jacobians of the disk averaged spectra for (top) \ce{CH3CN} and (bottom) \ce{CH3C^{15}N}, which express the sensitivity of spectral radiance to the volume mixing ratios at each altitude.
(b) Close up of an individual line ($J$=19(3)--18(3)) at 349.393 GHz and 338.882 GHz for (top) \ce{CH3CN} and (bottom) \ce{CH3C^{15}N}, respectively. 
}}
\label{fig:jacobians}
\end{center}
\end{figure}

\edit1{Contrary to the model proposed by \cite{Dobrijevic2018}, suggests the presence of a large gradient in the \ce{^{14}N/^{15}N} isotopic ratios with altitude, we have assumed a vertically constant isotopic ratio in this study. 
The quality of the \ce{CH3C^15N} spectrum used in this study is not sufficient to make it possible to retrieve vertically resolved information. 
This is a caveat of the current analysis.}\\
\edit1{ In order to check the information on which altitude the observed spectra have sensitivity, Jacobians for \ce{CH3CN} and \ce{CH3C^{15}N} are shown in Figure \ref{fig:jacobians}. 
A Jacobian is a matrix of partial derivatives of the spectral radiance at each frequency with respect to the volume mixing ratios of \ce{CH3CN} at each altitude, which express the altitude sensitivity to the species. 
The plotted Jacobians were calculated using the retrieved \ce{CH3CN} profile (as shown in Figure \ref{fig:CH3CN}(b)) and the best-fit \ce{^{14}N/^{15}N} value. 
It is shown that the emission at the line centers of \ce{CH3CN} has sensitivity to \ce{CH3CN} at altitudes of 160 -- $\sim$400 km with the peak weight located at $\sim$300 km. 
The probing altitude shifts downward with respect to the frequency from the line center.
The Jacobian of \ce{CH3C^{15}N} shows weaker sensitivity at a narrower altitude range compared to that of \ce{CH3CN}, which is due to the weaker line opacity of \ce{CH3C^{15}N}. 
Because we used the \ce{CH3CN} spectra not only the line center but also the line wings for the retrieval analysis, our derived \ce{CH3CN} profile is also sensitive to the lower altitudes to which the \ce{CH3C^{15}N} line has the sensitivity. 
In other words, the sensitive altitude ranges of \ce{CH3CN} and \ce{CH3C^{15}N} are not completely separated in this study.}

\subsection{Impact of the assumed temperature profile}
One of the potential error sources in this study is the uncertainty in the temperature profile used in the radiative transfer calculation. 
We tested different temperature profiles for the \ce{^{14}N/^{15}N} derivation. 
\cite{Thelen2018} derived not only a disk-averaged temperature profile, but also temperature profiles representing the northern and southern hemispheres of Titan (Figure \ref{fig:temperature}). 
Considering the fact that \ce{CH3CN} is enhanced in the northern hemisphere, as shown in Figure \ref{fig:CH3CN_map}, the use of the northern hemispheric temperature profile provides an alternative choice when selecting a temperature profile for analyzing the \ce{CH3CN} spectra. 
If we use the northern hemispheric temperature profile of \cite{Thelen2018}, the best fit \ce{^{14}N/^{15}N} in \ce{CH3CN} was \edit1{129, which is not largely different from that of using disk-averaged temperature profile}. 
\edit1{Thus, these results allow us to conclude that the uncertainty that results due to the temperature profile used in this study is not fatal. } 

\subsection{Comparisons with the model and future perspectives}
The best-fit \ce{^{14}N/^{15}N} value for \ce{CH3CN}, which is \edit1{125}, falls within the range of the values obtained in the theoretical study of \cite{Dobrijevic2018} (derived as 70--170 at 200 km, as the result of the Monte-Carlo simulation). 
As mentioned in Section 1, \ce{CH3CN} is expected to have two possible \ce{N}-atom origins, UV and GCR induced dissociation. 
\cite{Dobrijevic2018} simulated the contribution of these two processes to \ce{^{14}N/^{15}N} for \ce{CH3CN}, and found that \ce{^{14}N/^{15}N} has two possible peak values, 90 and 160 (at 200 km), which corresponds to the UV and GCR origin scenarios, respectively. 
However, due to the data quality limitation, the \ce{^{14}N/^{15}N} value derived in this study is not sufficiently accurate to clearly differentiate between these two possible \ce{^{14}N/^{15}N} scenarios. 
In contrast to \ce{CH3CN}, both \ce{HCN} and \ce{HC3N} were expected to exhibit low \ce{^{14}N/^{15}N} value as 79.5$\pm$6.5 and 79.6$\pm$6.5, respectively, in the model. 
Our derived best-fit value of \ce{^{14}N/^{15}N} for \ce{CH3CN} is inconsistent with that expected for both \ce{HCN} and \ce{HC3N}, which supports the model's prediction of the different \ce{N2} dissociation sources among these nitriles. 

For the further observational constraints, a dedicated observation of \ce{CH3CN}/\ce{CH3C^{15}N} with better sensitivity is required. 
In addition, derivation of a vertical profile of \ce{^{14}N/^{15}N} for \ce{CH3CN} could provide a good evaluation of the model because \cite{Dobrijevic2018} predicted that the vertical distribution of \ce{^{14}N/^{15}N} might possibly exhibit a large gradient. 
Additionally, both the temporal and spatial variations of \ce{^{14}N/^{15}N} derived by additional observations will be useful for facilitating better understanding of the contributions of production processes under different environments (different UV condition, for example).

\acknowledgments
This study makes use of the following ALMA data: ADS/JAO.ALMA$\#$2013.1.00033.S. 
ALMA is a partnership of ESO (representing its member states), NSF (USA) and NINS (Japan), together with NRC (Canada), MOST and ASIAA (Taiwan), and KASI (Republic of Korea), in cooperation with the Republic of Chile. The Joint ALMA Observatory is operated by ESO, AUI/NRAO and NAOJ. 
Authors are grateful to Koichiro Nakanishi and Hiroshi Nagai for their valuable comments and useful website for ALMA data reduction.  
TI wishes to thank Mitaro Namiki of the Tokyo University of Agriculture and Technology and staff of the Information Technology Center of the University of Tokyo for their dedicated work and suggestions for the computing and understanding of the entire research works.
\edit1{Authors are grateful to an anonymous reviewer who gave important comments on manuscript from the professional point of view.}
This work was financially supported by a Telecommunications Advancement Foundation(TI), the Japan Society for the Promotion of Science (JSPS) Kakenhi grant (17K14420 and 19K14782, TI) and an Astrobiology Center Program of National Institutes of Natural Sciences (NINS) grant.

\newpage

\end{document}